\documentclass[sigconf]{acmart}

\AtBeginDocument{%
  \providecommand\BibTeX{{%
    \normalfont B\kern-0.5em{\scshape i\kern-0.25em b}\kern-0.8em\TeX}}}

\copyrightyear{2023}
\acmYear{2023}
\setcopyright{acmlicensed}\acmConference[CIKM '23]{Proceedings of the 32nd
ACM International Conference on Information and Knowledge
Management}{October 21--25, 2023}{Birmingham, United Kingdom}
\acmBooktitle{Proceedings of the 32nd ACM International Conference on
Information and Knowledge Management (CIKM '23), October 21--25, 2023,
Birmingham, United Kingdom}
\acmPrice{15.00}
\acmDOI{10.1145/3583780.3615223}
\acmISBN{979-8-4007-0124-5/23/10}

\usepackage{graphicx}
\usepackage{eucal}
\usepackage{amsmath,scalerel}
\usepackage{multicol}
\usepackage{multirow}
\usepackage{booktabs}

\usepackage{amssymb}
\usepackage{caption}
\usepackage{subcaption}
\usepackage{bm}
\usepackage[ruled,linesnumbered]{algorithm2e}
\usepackage{color}

\begin{document}

\title{MI-DPG: Decomposable Parameter Generation Network Based on Mutual Information for Multi-Scenario Recommendation}

\author{Wenzhuo Cheng}
\orcid{0000-0001-7076-5779}
\affiliation{%
  \institution{Ant Group}
  \city{Hangzhou}
  \country{China}}
\email{wenzhuo.cwz@antgroup.com}

\author{Ke Ding}
\orcid{0009-0001-0562-1987}
\affiliation{%
  \institution{Ant Group}
  \city{Hangzhou}
  \country{China}}
\email{dingke.dk@antgroup.com}

\author{Xin Dong}
\orcid{0009-0004-2523-9971}
\affiliation{%
  \institution{Ant Group}
  \city{Shanghai}
  \country{China}}
\email{zhaoxin.dx@antgroup.com}

\author{Yong He}
\orcid{0009-0000-5390-2655}
\affiliation{%
  \institution{Ant Group}
  \city{Hangzhou}
  \country{China}}
\email{heyong.h@antgroup.com}

\author{Liang Zhang}
\orcid{0000-0002-7744-7789}
\affiliation{%
  \institution{Ant Group}
  \city{Shanghai}
  \country{China}}
\email{zhuyue.zl@antgroup.com}

\author{Linjian Mo}
\orcid{0000-0002-6682-1448}
\authornote{corresponding author}
\affiliation{%
  \institution{Ant Group}
  \city{Shanghai}
  \country{China}}
\email{linyi01@antgroup.com}

\renewcommand{\shortauthors}{Wenzhuo Cheng et al.}
\renewcommand{\shorttitle}{MI-DPG: Decomposable Parameter Generation Network Based on Mutual Information}

\begin{abstract}
Conversion rate (CVR) prediction models play a vital role in recommendation systems. Recent research shows that learning a unified model to serve multiple scenarios is effective for improving overall performance. However, it remains challenging to improve model prediction performance across scenarios at low model parameter cost, and current solutions are hard to robustly model multi-scenario diversity. In this paper, we propose MI-DPG for the multi-scenario CVR prediction, which learns scenario-conditioned dynamic model parameters for each scenario in a more efficient and effective manner. Specifically, we introduce an auxiliary network to generate scenario-conditioned dynamic weighting matrices, which are obtained by combining decomposed scenario-specific and scenario-shared low-rank matrices with parameter efficiency. For each scenario, weighting the backbone model parameters by the weighting matrix helps to specialize the model parameters for different scenarios. It can not only modulate the complete parameter space of the backbone model but also improve the model effectiveness. Furthermore, we design a mutual information regularization to enhance the diversity of model parameters across scenarios by maximizing the mutual information between the scenario-aware input and the scenario-conditioned dynamic weighting matrix. Experiments from three real-world datasets show that MI-DPG outperforms previous multi-scenario recommendation models.
\end{abstract}





\maketitle

\section{Introduction}\label{sec:Introduction}
 CVR prediction is an important technology in recommendation systems. These models usually need to serve multiple different scenarios in the system. However, it is difficult to design a unique prediction model for each scenario due to the limitation of computing resources. Therefore, adopting a unified modeling framework to integrate data from multiple scenarios is a common solution, called multi-scenario recommendation.

Recent research on multi-scenario recommendation can be divided into two categories~\cite{Yang2022AdaSparseLA}. (1) Models with individual parameters~\cite{tang2020progressive,sheng2021one}, such studies manually maintain individual learnable parameters for each scenario, which is inefficient in model parameter cost. (2) Models with generated parameters~\cite{Yang2022AdaSparseLA,Chang2023PEPNetPA,yan2022apg}, such studies use scenario-aware features as input to the auxiliary network and dynamically generates scenario-specific model layer parameters or output gating. ~\cite{Yang2022AdaSparseLA,Chang2023PEPNetPA} dynamically generate 
the output gating of the model layer. Due to the limited impact on the complete parameter space of the model layer, these are less effective. ~\cite{yan2022apg} dynamically generates the  model layer parameters for each scenario. Although low-rank decomposition of the layer parameters reduces the model parameter cost, the model performance decreases sharply with decreasing low-rank matrix size. It is difficult to achieve high cross-scenario prediction performance at a low model cost. Meanwhile, the parameter distribution generated by directly applying the auxiliary network suffers from insufficient diversity, and cannot robustly model the diversity of multiple scenarios. There are two crucial issues: (1) How to obtain more effective cross-scenario prediction performance at a more efficient model parameter cost? (2) How to model multi-scenario diversity more robustly?

In this paper, we propose a framework MI-DPG, which models multi-scenario diversity in a more efficient and effective manner. Specifically, we introduce a scenario-conditioned dynamic weighting matrix to measure the importance of model parameters under different scenarios, which is obtained by combining the decomposed scenario-specific and scenario-shared low-rank matrices. Our model can be more efficiently reparameterized for different scenarios, resulting in more effective cross-scenario prediction performance. In particular, the performance of our model does not decrease as the size of the low-rank matrix decreases, since we do not directly perform low-rank decomposition on the backbone model layer. In addition, we propose a mutual information regularization to increase the correlation between the scenario-aware input and the dynamic weighting matrix by maximizing their mutual information, thereby enhancing the diversity of model parameters across scenarios~\cite{Chen2016InfoGANIR,Hjelm2018LearningDR,Belghazi2018MINEMI}. In summary, our contributions are:
\vspace{-\topsep}
\begin{itemize}
    \item We propose MI-DPG to learn dynamic model parameters weighted by a decomposable weighting matrix for each scenario, which can be more efficiently reparameterized for each scenario while achieving more effective performance.
    \item We propose mutual information regularization to maximize the mutual information between the scenario-aware input and the scenario-conditioned dynamic weighting matrix to increase their correlation, thereby more robustly modeling multi-scenario diversity.
    \item Experimental results from three real-world datasets demonstrate that our proposed model outperforms state-of-the-art methods in terms of AUC and LogLoss metrics.
\end{itemize}
\vspace{-\topsep}

\begin{figure}[htb]
\centering
\includegraphics[width=8cm]{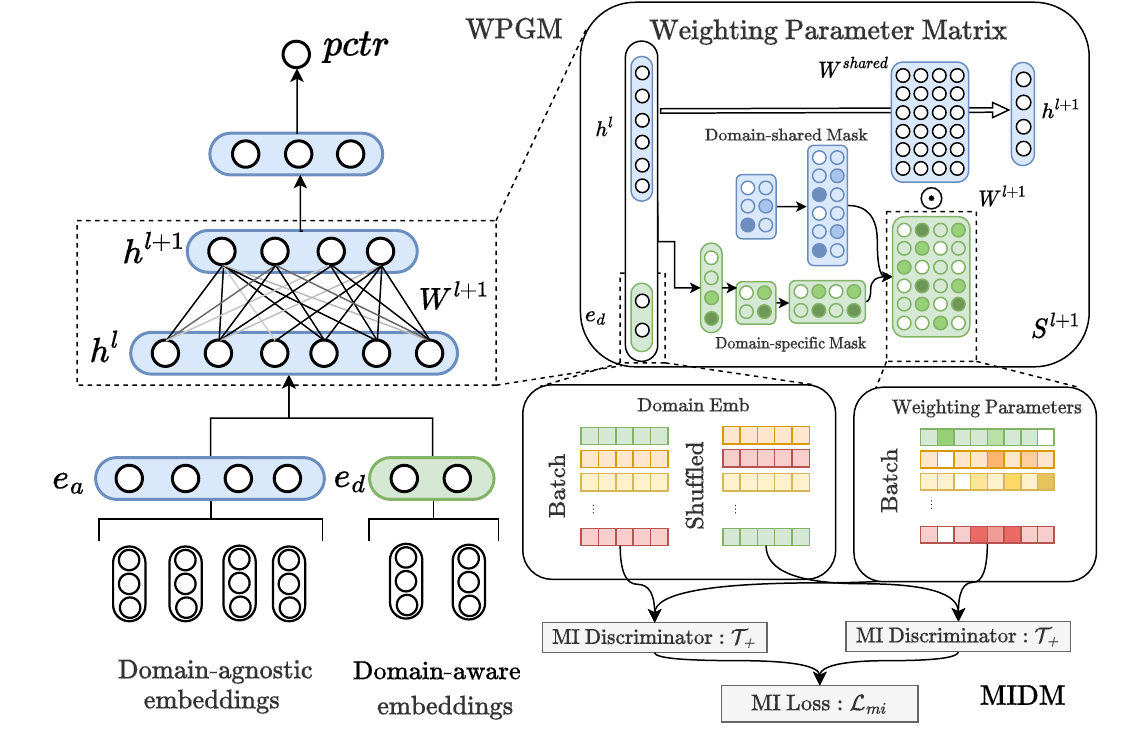}
\caption{MI-DPG for multi-scenario recommendation.}
\label{fig:figure_1}
\end{figure}

\section{Methodology}\label{sec:method}
In this section, we describe the framework MI-DPG in detail. As shown in Figure~\ref{fig:figure_1}, the core of our method has two parts. The Decomposable Weighting Matrix Module (DWM) learns dynamic model parameters. The Mutual Information Regularization Module (MIR) optimizes the mutual information of model parameters.

\subsection{Preliminaries}\label{sec:Preliminaries}
The training dataset can be denoted as $\mathcal{D} = \left \{ \left ( x_{i},y_{i}\right )  \right \} _{i=1}^{\left | \mathcal{D}  \right | }$, where $x_{i}$ and $y_{i}$ denote the feature set and binary conversion label of $i_{th}$ sample respectively. Since we mainly discuss research related to the multi-scenario recommendation, the dataset $\mathcal{D}$ can be further divided into multiple scenario-specific subsets $\mathcal{D}$ ($\mathcal{D}=\bigcup \mathcal{D}^{d}$). A subset of scenario $d$ can be expressed as $\mathcal{D}^{d}=\left \{ \left ( x_{i}^{ag},x_{i}^{aw},y_{i} \right )  \right \} _{i=1}^{\left | \mathcal{D}^{d} \right | }$, where $x_{i}^{aw}$ is the scenario-aware feature set and $x_{i}^{ag}$ is the scenario-agnostic feature set. After transforming all features into embeddings, we concatenate scenario-agnostic and scenario-aware feature embeddings as $\left [ e_{i}^{aw},e_{i}^{ag} \right ]$, which is used as the backbone model input. At the same time, we use an L-layered fully-connected neural network as the backbone model architecture:
\begin{small}
\begin{align}
\tilde{y_{i}}= \left \{ \mathcal{F}_{S_{i}^{l}\odot W^{l}} \left ( h^{l}  \right ) , S_{i}^{l}=DWM(e_{i}^{aw},h^{l})\right \}_{1\le l\le L}
\end{align}
\end{small}

where $\odot $ is element-wise multiplication. $\mathcal{F}$ is the backbone neural network. $W^{l} \in \mathbb{R} ^{d_{m} \times d_{n} }$ is the parameter matrix of the backbone network. $S_{i}^{l} \in \mathbb{R}^{d_{m} \times d_{n} }$ is the weighting matrix dynamically generated by the \textit{DWM} (see Section~\ref{sec:DWM} for details) that takes scenario-aware embeddings $e_{i}^{aw}$ and $h^{l}$ of $i_{th}$ sample, where $h^{l}$ is the input of the $l_{th}$ layer of the backbone model. When $l=1$, $h^{l}$ is $\left [ e_{i}^{aw},e_{i}^{ag} \right ]$.

\subsection{Decomposable Weighting Matrix Module}\label{sec:DWM}
We propose a variety of processing methods according to whether to decompose the matrix and how to decompose the matrix:

\subsubsection{\textbf{Non-decomposable(Non)}}\label{sec:Non}
This way is to directly generate the undecomposed scenario-conditioned weighting matrix through the auxiliary network $\mathcal{G}$, where $\mathcal{G}$ is a lightweight fully-connected neural network. The formula can be defined as:
\begin{small}
\begin{align}
S_{i}^{l} = \sigma (\mathcal{G}(e_{i}^{aw},h^{l}))
\end{align}
\end{small}
where $S_{i}^{l} \in \mathbb{R} ^{d_{m} \times d_{n} }$, $\sigma$ is the sigmoid activation function.

\subsubsection{\textbf{Row-wise Specific and Column-wise Specific(RC)}}\label{sec:RC}
Inspired by low-rank decomposition methods~\cite{Hu2021LoRALA,Wen2017CoordinatingFF,Kalman1996ASV} which have demonstrated that robust performance can be achieved by optimizing tasks in low-rank subspaces, we hypothesize that the resulting dynamic weighting matrix also have low intrinsic rank. So we propose to decompose the weighting matrix $S_{i}^{l} \in \mathbb{R} ^{d_{m} \times d_{n} }$ as two low-rank sub-matrices, which are $S_{R}^{l} \in \mathbb{R} ^{d_{r} \times d_{k} }$ and $S_{C}^{l} \in \mathbb{R} ^{d_{k} \times d_{c} }$. Note that $d_{r}$, $d_{k}$ and $d_{c}$ are user-specified hyperparameters that control the size of the low-rank matrix. For simplicity, we consider $d_{r}<=d_{m},d_{m} \bmod d_{r}=0$, and $d_{c}<=d_{n},d_{n} \bmod d_{c}=0$, and $d_{k}\ll min(d_{m},d_{n})$. The formula can be defined as:
\begin{small}
\begin{align}
S_{i}^{l} = \psi (\sigma (S_{R}^{l}S_{C}^{l})),
S_{R}^{l}=\mathcal{G}_{R}(e_{i}^{aw},h^{l}),S_{C}^{l}=\mathcal{G}_{C}(e_{i}^{aw},h^{l})
\end{align}
\end{small}
where $S_{i}^{l}$ is generated by the outer product of $S_{R}^{l}$ and $S_{C}^{l}$. $\psi$ is a repeat function that repeats its input $\frac{d_{m}}{d_{r}}$ times on the row axis and $\frac{d_{n}}{d_{c}}$ times on the column axis. When $d_{r}=1,d_{k}=1$ and $d_{c}=d_{n}$, our method is equivalent to methods ~\cite{Yang2022AdaSparseLA,Chang2023PEPNetPA}. It shows that our method is more general and can modulate the parameter space of the whole backbone model. In addition, unlike the direct matrix factorization of the backbone model parameters in method~\cite{yan2022apg}, we perform matrix factorization on the weighting matrix, which can achieve higher performance at a lower low-rank matrix size.

\subsubsection{\textbf{Row-wise Specific and Column-wise Shared(R)}}\label{sec:R}
First, we hypothesize that it is beneficial to share one of the two low-rank matrices as cross-scenario, which enables the dynamic weighting matrix to generalize better across scenarios. Then, by keeping the column-wise low-rank matrix as global. This can be expressed as:
\begin{small}
\begin{align}
S_{i}^{l} = \psi (\sigma (S_{R}^{l}C_{share}^{l})),S_{R}^{l}=\mathcal{G}_{R}(e_{i}^{aw},h^{l})
\end{align}
\end{small}
where $S_{i}^{l} \in \mathbb{R} ^{d_{m} \times d_{n} }$, which is generated by the outer product of $S_{R}^{l} \in \mathbb{R} ^{d_{r} \times d_{k} }$ and $C_{share}^{l} \in \mathbb{R} ^{d_{k} \times d_{c} }$. Hence, the outer product is essentially a rich dyadic composition between scenario-specific and scenario-shared low-rank matrices. 

\subsubsection{\textbf{Row-wise Shared and Column-wise Specific(C)}}\label{sec:C}
Here, we reverse the row-wise setting as scenario-shared low-rank matrices, and the other settings are ditto. This can be expressed as:
\begin{small}
\begin{align}
S_{i}^{l} = \psi (\sigma (R_{share}^{l}S_{C}^{l})),S_{C}^{l}=\mathcal{G}_{C}(e_{i}^{aw},h^{l})
\end{align}
\end{small}

\subsection{Mutual Information Regularization Module}\label{sec:MIR}
Although the above methods can learn scenario-conditioned dynamic model parameters for each scenario, we find that directly applying the auxiliary network leads to insufficient diversity in the dynamic weighting matrix distributions. We hope that input instances belonging to different scenarios generate different dynamic weighting matrix distributions, while input instances belonging to the same scenario generate similar distributions. To this end, we propose the mutual information regularization, which is to maximize the mutual information between the scenario-aware input and the generated dynamic weighting matrix. According to the research~\cite{Asano2008ENTROPYR}, we know that mutual information measures the degree of interdependence between two variables. The more relevant the dynamic weighting matrix is to the scenario-aware input, the more its distribution can show the differences between multiple scenarios. The formula of the MIR module is described as follows:
\begin{small}
\begin{align}
    \mathcal{L}_{mi}=-\frac{1}{n} \sum_{i=1}^{N}\left [\log_{}{\mathcal{T_{+}}}+\log_{}{(1-\mathcal{T_{-}})} \right ] \\
    \mathcal{T_{+}}=\mathcal{T}(e_{i}^{aw},S_{i}^{l}),\mathcal{T_{-}}=\mathcal{T}(shuffle(e_{j}^{aw}),S_{i}^{l})
\end{align}
\end{small}
where $N$ is the number of training samples. $e_{i}^{aw}$ is the scenario-aware input, and $S_{i}^{l}$ is the dynamic weighting matrix. $shuffle$ is a random shuffling of $e_{i}^{aw}$ within the training samples. $\mathcal{T}$ is the divergence that measures the relatedness of the two distributions. Inspired by mutual information maximization estimation~\cite{Hjelm2018LearningDR,Nowozin2016fGANTG}, $\mathcal{T_{+}}$ encourages the dynamic weighting matrix distribution to be more relevant to the corresponding scenario, while $\mathcal{T_{-}}$ encourages the dynamic weighting matrix distribution to be more irrelevant to the non-corresponding scenario, thus making the distribution of the dynamic weighting matrix in each scenario more different. In our implementation~\cite{Nowozin2016fGANTG}, $\mathcal{T_{+}}$ and $\mathcal{T_{-}}$ are a parameter-shared discriminator module, implemented by a lightweight neural network, which can be used to estimate the divergence $\mathcal{T}$.

Finally, the overall optimization goal of our model MI-DPG is:
\begin{small}
\begin{align}
    \mathcal{L}_{CVR}=-\frac{1}{n} \sum_{i=1}^{N}y_{i}\log_{}{\tilde{y_{i}}}+ \lambda \mathcal{L}_{mi}
\end{align}
\end{small}
the first item is the cross-entropy loss function of the conversion behavior of the CVR model, $\mathcal{L}_{mi}$ is the optimization target of \textit{MIR} module, and $\lambda$ are hyperparameters.

\section{Experiments}
In order to verify the effectiveness of the proposed  MI-DPG, we evaluate and compare the performance of MI-DPG and prior methods on three real-world datasets.

\subsection{Datasets}\label{sec:datasets}
\subsubsection{Amazon Dataset (Public).}\label{sec:Amazon}
This public dataset is collected from the electronics category on Amazon. which contains 1,292,954 instances, 1,157,633 users and 10 categories. We will abbreviate the category name as follows: Television $\&$ Video (TV), Security $\&$ Surveillance (SS), Car Electronics $\&$ GPS (CG), Wearable Technology (WT), Computers $\&$ Accessories (CA), Camera $\&$ Photo ( CP), Home Audio(HA), Portable Audio $\&$ Video(PV), Headphones(HP), Accessories $\&$ Supplies(AS).
\subsubsection{IAAC Dataset (Public).}
This public dataset is a dataset collected from sponsored searches in e-commerce. Each record refers to whether the user purchased the displayed product after clicking the product. There are 478,138 records, 197,694 users and 10,075 items.
\subsubsection{Advertising Dataset (Industrial).}
We collected user traffic logs from the commercial advertising system of the online application. Each record refers to whether a conversion behavior occurred after the user clicked on the product. The dataset includes 108,649,408 click samples from 43,397,838 users, which can be further subdivided into: 3 scenarios, 10 trades, 300 advertisers, 6,203 groups and 13,571 ads.

\subsection{Baseline Models}\label{sec:baseline models}
We use DNN~\cite{Covington2016DeepNN} as the backbone of the CVR prediction model. First, we compare MI-DPG with two classes of multi-scenario recommendation methods introduced in Section~\ref{sec:Introduction}. (1) PLE~\cite{tang2020progressive} learns predefined scenario-specific experts using the data from this scenario only. (2) STAR~\cite{sheng2021one} learns predefined scenario-specific parameters using the data from this scenario only. (3) PEPNet~\cite{Chang2023PEPNetPA} uses an auxiliary network to generate output gating. (4) AdaSparse~\cite{Yang2022AdaSparseLA} uses an auxiliary network to generate neuron-level weighting factors to prune the model structure. (5) APG~\cite{yan2022apg} uses auxiliary networks to generate scenario-specific parameters, and performs low-rank decomposition of layer parameters to reduce model parameter costs. Then, we also compare the variant model called Decomposable Parameter Generation Network (DPG) of MI-DPG, which does not have the MIR module.

\subsection{Experimental Setup}
The metrics of \textit{AUC}~\cite{Fawcett2006AnIT} and \textit{LogLoss} are used for evaluation. All models are implemented in Tensorflow version 1.13 using the Adam optimizer~\cite{Kingma2014AdamAM} with an initial learning rate of 0.001. The mini-batch size is fixed at 512. The backbone of the network is a \textit{DNN}~\cite{Covington2016DeepNN} with three hidden layers of dimensions 128, 64, 32. For each compared model, we carefully tune their hyperparameters. The hyperparameters in \textit{MI-DPG} are set as follows: $\lambda=1$. Also, we run 10 trials per experiment for more stable results.

\subsection{Experimental Results}\label{sec:Experimental Results}
Table~\ref{tab:table1} shows the comparison results of multi-scenario CVR prediction methods. First, we noticed that PEPNet, AdaSparse, and APG are superior to PLE and STAR, and verified that models with generated parameters are superior to models with individual parameters. Second, DPG outperforms the above models, indicating that our DWM module can achieve more effective cross-scenario prediction performance at a more efficient model parameter cost. Third, MI-DPG achieves the highest AUC and the lowest LogLoss on all datasets, indicating that our MIR module further improves model performance by modeling multi-scenario diversity more robustly.

\renewcommand\arraystretch{1.2}
\begin{table}[!htbp]
\footnotesize
\centering
\caption{Performance comparison on multi-scenario CVR prediction. "*" indicates the statistical significance for $p <= 0.05$ compared with the best baseline
 based on the paired t-test.}
\scalebox{1}{
\begin{tabular}{c|cc|cc|cc}
\toprule
\multirow{2}{*}{\textbf{Method}}  & \multicolumn{2}{c}{\textbf{Amazon}} & \multicolumn{2}{c}{\textbf{IAAC}}  &\multicolumn{2}{c}{\textbf{Advertising}}  \\ 
\cline{2-7} & AUC  & LogLoss & AUC   & LogLoss & AUC   & LogLoss \\ 
\midrule\midrule
DNN & 0.6590 & 0.5235 & 0.6335 & 0.0865 & 0.7443 & 0.2434  \\ 
\hline
PLE & 0.6639  & 0.5201  & 0.6377  & 0.0846  & 0.7464  & 0.2391  \\ 
STAR & 0.6623 & 0.5205 & 0.6424 & 0.0829 & 0.7447 & 0.2439  \\ 
PEPNet & 0.6671 & 0.5167 & 0.6435 & 0.0825 & 0.7476 & 0.2381   \\ 
AdaSparse & 0.6695 & 0.5142 & 0.6462 & 0.0810 & 0.7491 & 0.2350   \\ 
APG        & 0.6697 & 0.5138 & 0.6457 & 0.0817  & 0.7488  & 0.2356  \\
\hline
\textbf{DPG} & 0.6716 & 0.5104 & 0.6479 & 0.0801 & 0.7533  & 0.2297 \\
\textbf{MI-DPG} & \textbf{0.6745*} & \textbf{0.5062*}  & \textbf{0.6504*} & \textbf{0.0794*} & \textbf{0.7580*} & \textbf{0.2237*}  \\
\bottomrule
\end{tabular}
}
\label{tab:table1}
\end{table}
\vspace{-1em}

\subsection{Empirical Analysis}\label{sec:Empirical Analysis}
\subsubsection{\textbf{Effect of Decomposable Weighting Matrix Module}}\label{sec:Ablation Study}
In order to verify the effect of the DWM module, we analyze from two angles of different DWN module variants and different low-rank matrix sizes. Please note that the analysis here is based on DPG without mutual information regularization. For the first angle, we propose the following four DPG variants: DPG-non, DPG-rc, DPG-r, and DPG-c, the details of which correspond to Section~\ref{sec:Non}-~\ref{sec:C} respectively. The experimental results are shown in Figure(~\ref{fig:figure_2a})(~\ref{fig:figure_2c})(~\ref{fig:figure_2e}), comprehensively compare the best results of the four variants on the three datasets (points marked with "$\star$"). (1) DPG-rc, DPG-r and DPG-c are mostly better than DPG-non in this case, which verifies the effectiveness of the low-rank decomposition operation on the dynamic weighting matrix. (2) DPG-rc is better than DPG-c but worse than DPG-r, indicating that it is beneficial to use row-wise low-rank matrices as cross-scenario shared embeddings, which can generalize better across scenarios.

For the second angle, we adjust the low-rank matrix size by combining hyperparameters ($d_{k}\in \left \{1,2,4\right \}$, $d_{r}\in \left \{ \frac{d_m}{8},\frac{d_m}{4},\frac{d_m}{2},\frac{d_m}{1}\right \}$ and $d_{c}\in \left \{ \frac{d_n}{8},\frac{d_n}{4},\frac{d_n}{2},\frac{d_n}{1}\right \}$) with grid search. Here we assume that $d_r=\frac{d_m}{2} $, $d_c=\frac{d_n}{2} $ and $d_k=1$, then the size of the low-rank matrix can be recorded as $\frac{1}{4}$ ($\frac{1}{2} \times \frac{1}{2} \times 1$). Therefore, the smaller the size of the low-rank matrix, the lower the cost of the model parameters. At the same time, in order to compare with APG~\cite{yan2022apg}, we also adjust the low-rank decomposition hyperparameters of APG accordingly while ensuring that the parameters of the two models are aligned. The experimental results are shown in Figure(~\ref{fig:figure_2a})(~\ref{fig:figure_2c})(~\ref{fig:figure_2e}): (1) Compared with APG on the three datasets, benefiting from the fact that we do not directly perform low-rank decomposition on the backbone model layer, Our four DPG variants maintain high model performance at lower low-rank matrix sizes. (2) Our four DPG variants have stable model performance across different low-rank matrix sizes, and even have the best results at smaller low-rank matrix sizes. It is verified that our model can achieve more effective cross-scenario prediction performance at a more efficient model parameter cost.

\subsubsection{\textbf{Effect of Mutual Information Regularization Module}}\label{sec: Collapse Analysis}
\begin{figure}[htb]
     \centering
     \begin{subfigure}[b]{4.2cm}
         \centering
         \includegraphics[width=4.1cm]{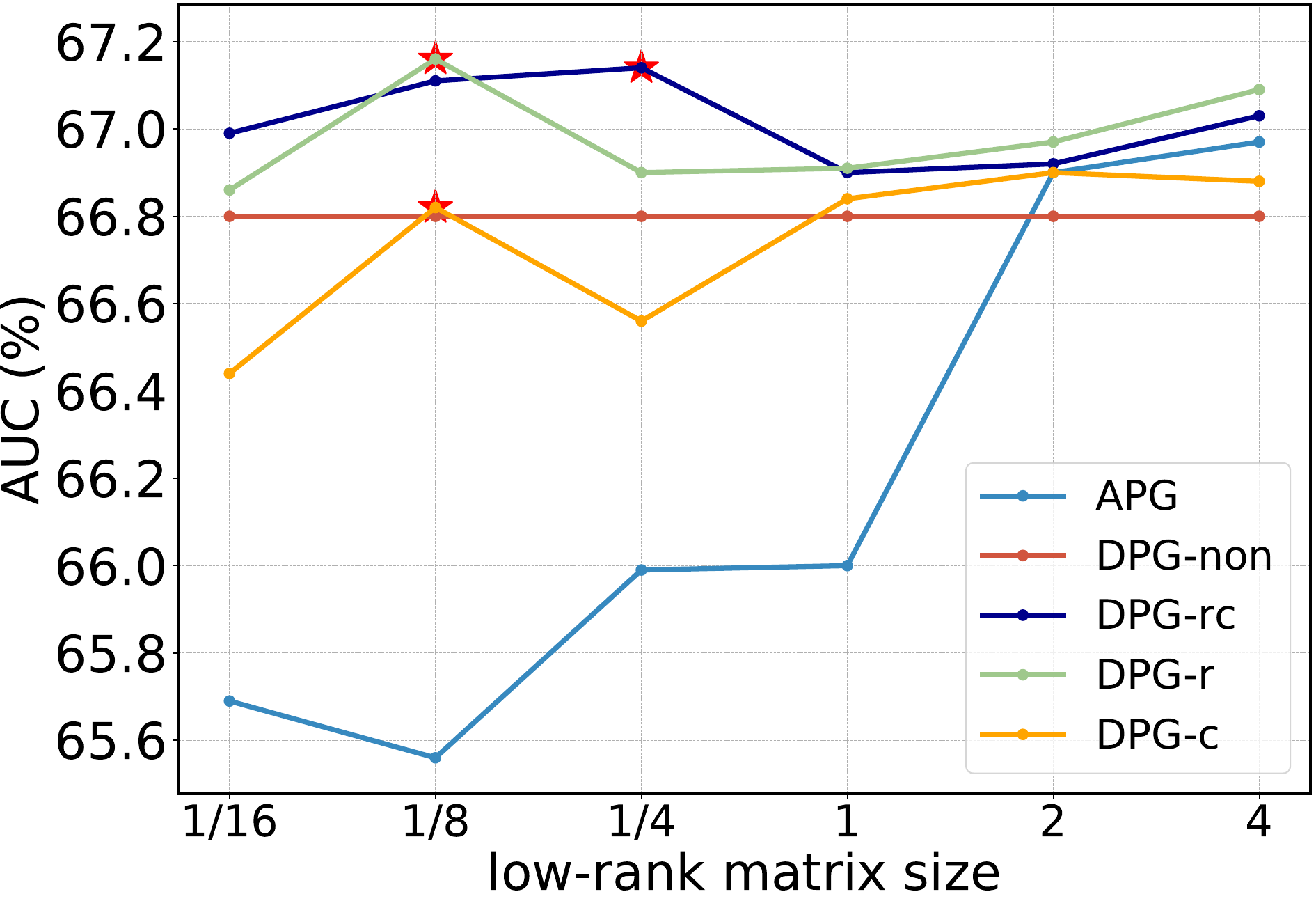}
         \captionsetup{font={scriptsize,stretch=1},justification=raggedright}
         \caption{AUC vs. low-rank size on Amazon.}
         \label{fig:figure_2a}
     \end{subfigure}
     \begin{subfigure}[b]{4.2cm}
         \centering
         \includegraphics[width=4.0cm]{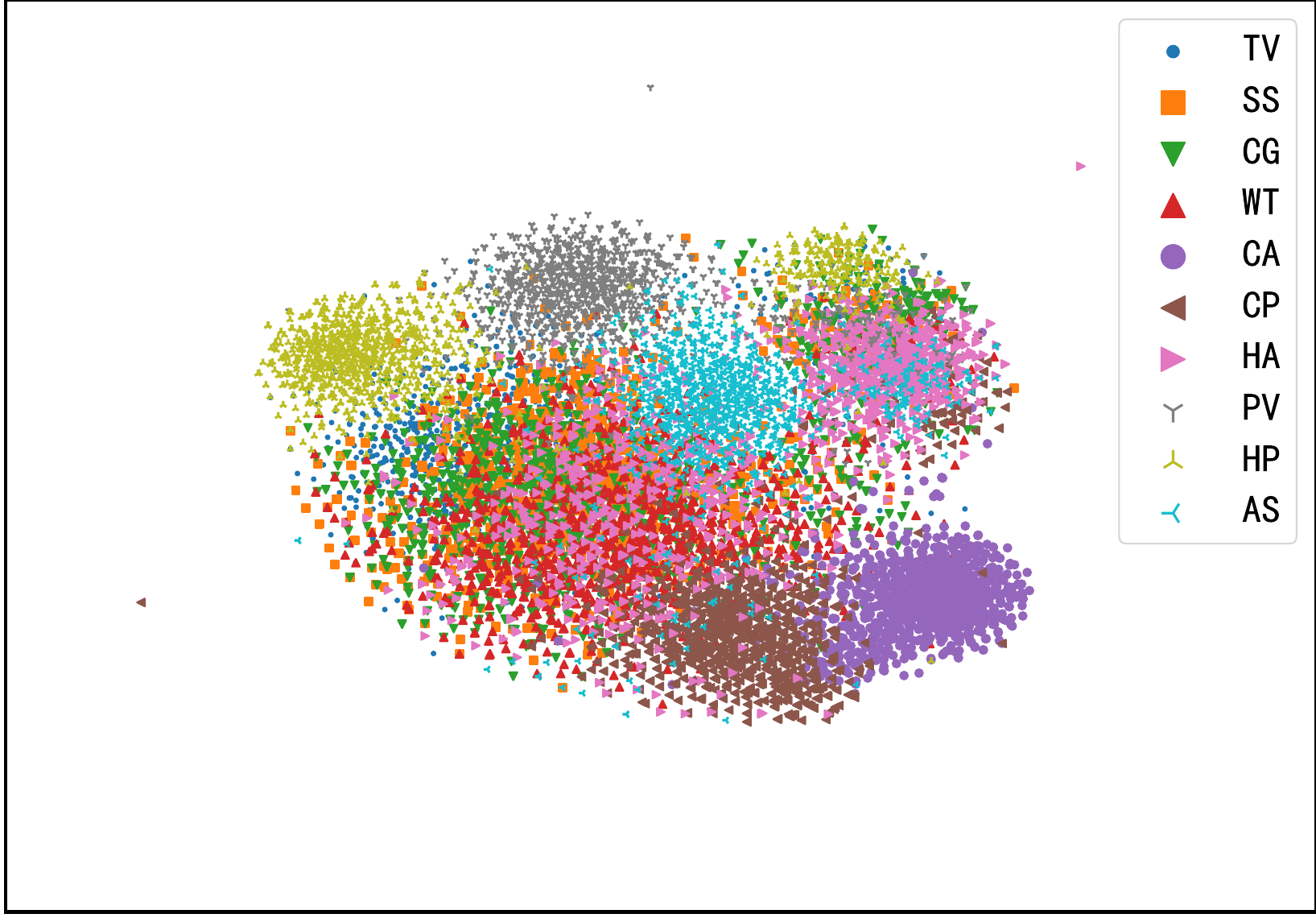}
         \captionsetup{font={scriptsize,stretch=1},justification=raggedright}
         \caption{Multi-scenario diversity $w/o$ $MIR$.} 
         \label{fig:figure_2b}
     \end{subfigure}
     \begin{subfigure}[b]{4.2cm}
         \centering
         \includegraphics[width=4.1cm]{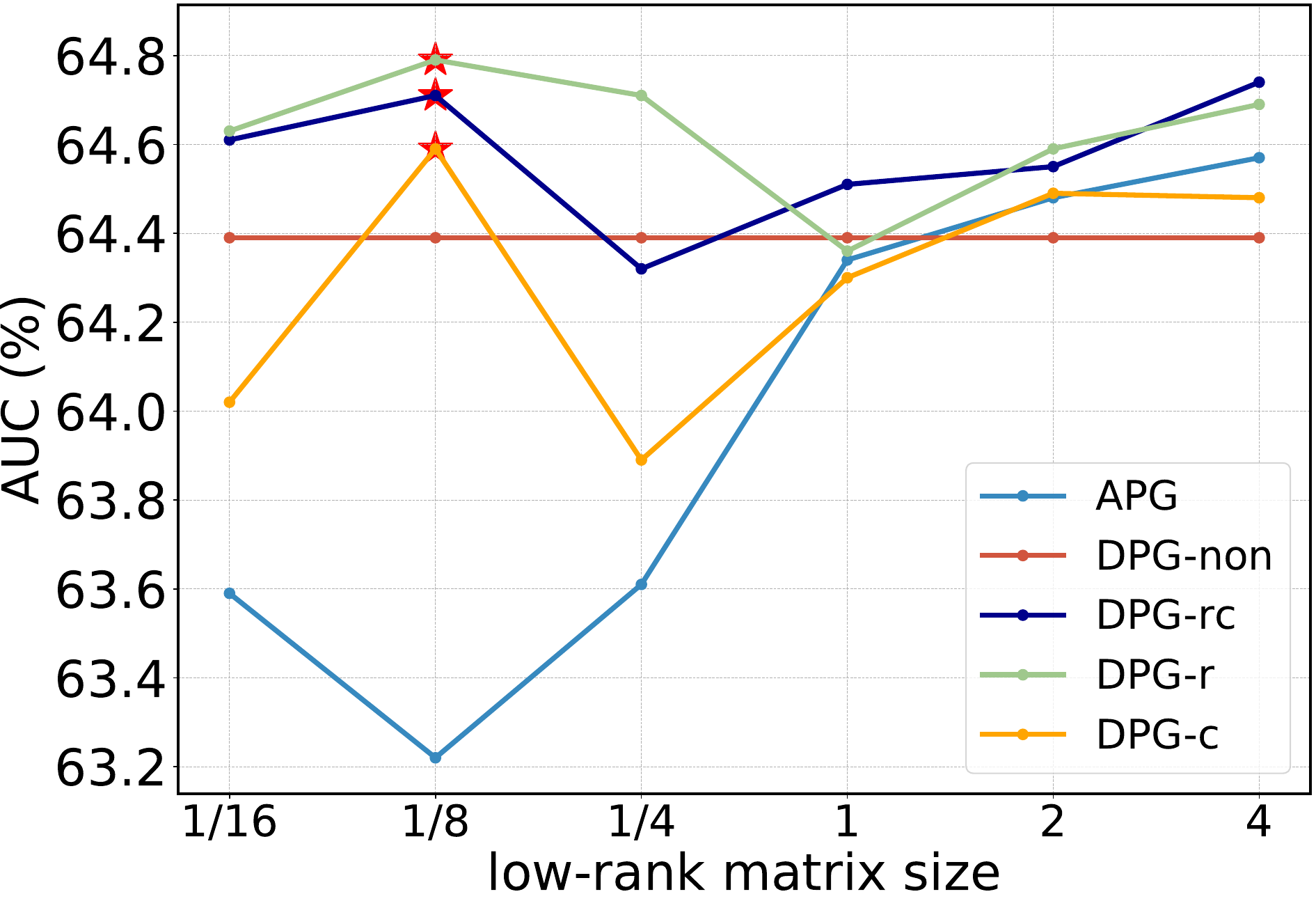}
         \captionsetup{font={scriptsize,stretch=1},justification=raggedright}
         \caption{AUC vs. low-rank size on IAAC.}
         \label{fig:figure_2c}
     \end{subfigure}
     \begin{subfigure}[b]{4.2cm}
         \centering
         \includegraphics[width=4.0cm]{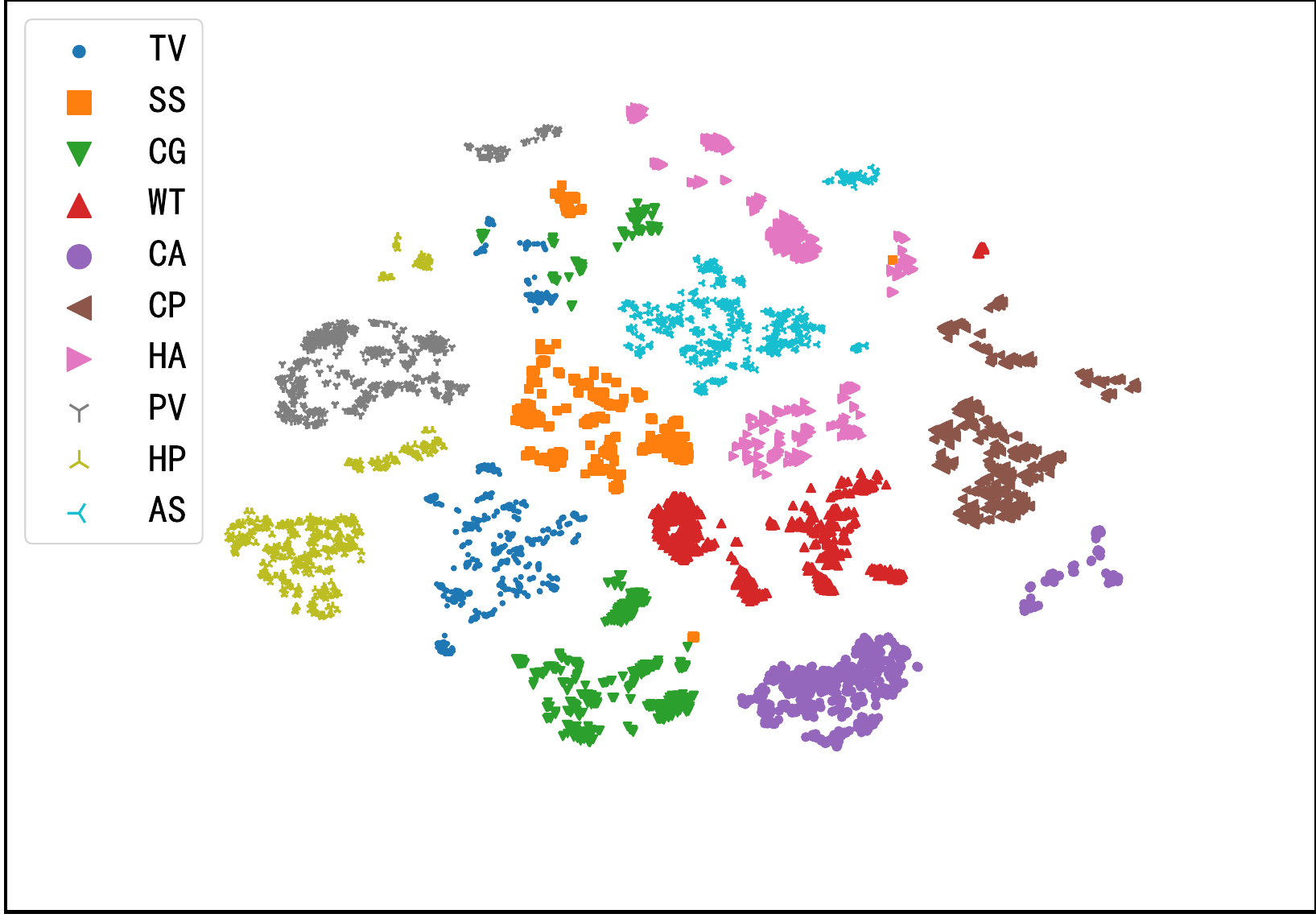}
         \captionsetup{font={scriptsize,stretch=1},justification=raggedright}
         \caption{Multi-scenario diversity $w/$ $MIR$.}
         \label{fig:figure_3d}
     \end{subfigure}
     \begin{subfigure}[b]{4.2cm}
         \centering
         \includegraphics[width=4.1cm]{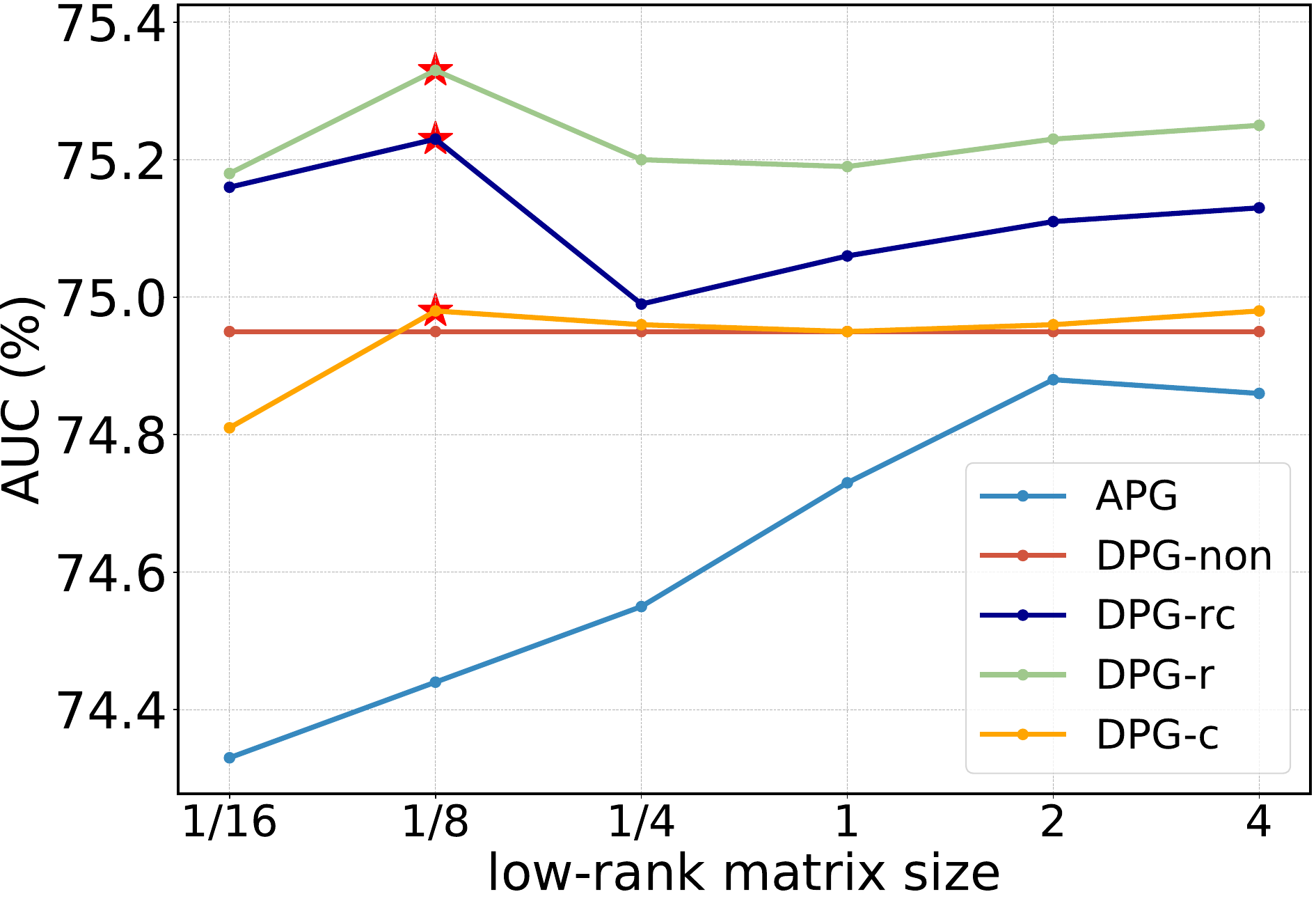}
         \captionsetup{font={scriptsize,stretch=1},justification=raggedright}
         \caption{AUC vs. low-rank size on Advertising.}
         \label{fig:figure_2e}
     \end{subfigure}
     \begin{subfigure}[b]{4.2cm}
         \centering
         \includegraphics[width=4.0cm]{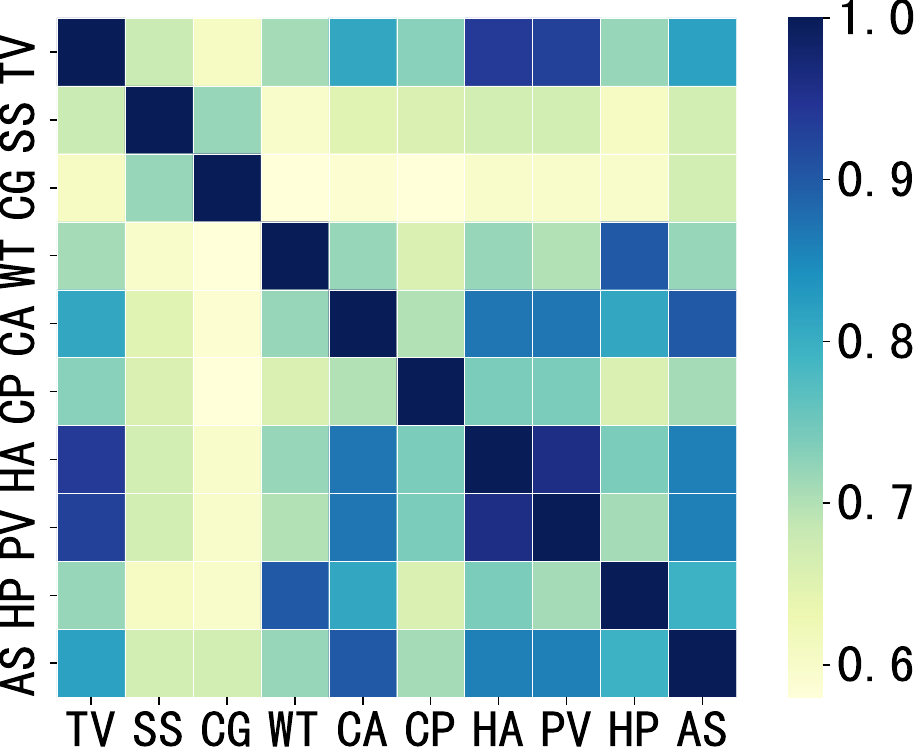}
         \captionsetup{font={scriptsize,stretch=1},justification=raggedright}
         \caption{Similarities of the weighting matrix.}
         \label{fig:figure_2f}
     \end{subfigure}
    \caption{Visualization of the effectiveness of DWM and MIR.}
    \label{fig:figure_3}
\end{figure}

In order to further determine whether the MIR module can enhance the diversity of dynamic weight matrix distributions in different scenarios. We use the item category on Amazon (see Section~\ref{sec:Amazon} for details of the category name) as the scenario-aware input, and visualize the generated dynamic weighting matrix through the t-SNE~\cite{Maaten2008VisualizingDU}. The experimental results are shown in Figure(~\ref{fig:figure_2b})(~\ref{fig:figure_3d}). The comparison between $w/$ MIR and $w/o$ MIR shows that mutual information regularization isolates the distribution of dynamic weighting matrices belonging to different scenarios, and at the same time shortens the distribution of dynamic weighting matrices belonging to the same scenario. This suggests that maximizing the mutual information between the scenario-aware input and the dynamic weighting matrix can enhance its diversity. At the same time, we compute the similarity between the means of the dynamic weighting matrix distributions belonging to the same scenario. As shown in Figure~\ref{fig:figure_2f}, the categories corresponding to the darker blocks (larger correlation) in the figure are all real and similar. This indicates that the scenario-conditioned dynamic model parameters learned by our model are meaningful and can more robustly model multi-scenario diversity.

\vspace{-1.0em}
\section{Conclusion}
We propose MI-DPG to learn dynamic model parameters weighted by a decomposable weighting matrix for each scenario, and optimize the mutual information of model parameters to model multi-scenario diversity in a more efficient and effective manner. Experimental results from three real-world datasets show that MI-DPG outperforms previous multi-scenario recommendation methods. 

\bibliographystyle{ACM-Reference-Format}
\bibliography{references}


\end{document}